\documentclass[runningheads,a4paper]{llncs}

\usepackage{url}

\usepackage{color}
\usepackage{alltt}
\usepackage[utf8]{inputenc}

\usepackage{graphicx}

\input {highlight.sty}

\usepackage{float}
\floatstyle{ruled}
\restylefloat{figure}
\usepackage{
  amsmath,
  amsfonts,
  amssymb,
  amsxtra,
  amsopn}
\usepackage{array}

\usepackage{tikz}
\usepackage[plain]{fancyref}

\newcommand{\BigO}[1]{\ensuremath{\operatorname{O}\bigl(#1\bigr)}}

\newcommand{\slice}[3]{

  \draw (0,0) -- (#1:1) arc (#1:#2:1);
  \draw (0,0) -- (#1:1.1);

  \node at (#1:1.3) {#3};
}

\newcommand{\segment}[3]{

  \draw (0,0) -- (#1:1) arc (#1:#2:1);
  \draw (0,0) -- (#1:1.1);

  \node at (#1:1.3) {#3};

  \pgfmathparse{0.5*#1+0.5*#2}
  \let\midangle\pgfmathresult
  \node at (\midangle:0.75) {#3};
}

\begin{document}

\mainmatter

\title{Perfect Consistent Hashing}

\author{Matthew Sackman\\\email{matthew@wellquite.org}}
\authorrunning{Matthew Sackman}
\institute{}
\maketitle

\begin{abstract}
Consistent Hashing functions are widely used for load balancing across
a variety of applications. However, the original presentation and
typical implementations of Consistent Hashing rely on randomised
allocation of hash codes to keys which results in a flawed and
approximately-uniform allocation of keys to hash codes. We analyse the
desired properties and present an algorithm that perfectly achieves
them without resorting to any random distributions. The algorithm is
simple and adds to our understanding of what is necessary to create a
consistent hash function.
\end{abstract}

\section{Introduction}

A hash function is a function that deterministically and uniformly
maps keys of unbounded size to members of a finite set of hash
codes. It is {\em deterministic} in that in the absence of changes to
the set of hash codes, the same key is always mapped to the same hash
code. It is {\em uniform} in that each hash code is equally likely to
be generated. If $h_C(\kappa)$ is the hash function $h$ with the set
of hash codes $C$, applied to the key $\kappa$, then $\forall \sigma
\in C, \forall \kappa.\: p(h_C(\kappa) = \sigma) = 1/|C|$.

A typical simple hash function is to treat the key as a natural number
and then to take its modulus by the number of hash codes, $|C|$. This
result is then used as an index into $C$, which is ordered in some way
and treated as mapping naturals to hash codes. If $C[x]$ represents
the result of indexing this mapping $C$ by $x$ then this simple hash
function can be written as $h_C(\kappa) = C[\kappa \mod |C|]$. One
problem with such a simple hash function is that when a new hash code
is added to $C$ or an existing hash code removed, the existing mapping
between keys and hash codes is non-minimally altered. If $C \subset
C'$ and $|C'| = |C| + 1$ (so $|C|$ and $|C'|$ are relatively prime)
then we can define the set of keys which do not get remapped as
\[ \{\kappa \mid \delta \in \{0 \dots (|C| - 1)\},\, \iota \in
\mathbb{N},\, \kappa = \delta + \iota \cdot |C| \cdot |C'|\} \] Thus
in any set of keys of size $|C| \cdot |C'|$, there will be on average
only $|C|$ keys for which $h_C(\kappa) = h_{C'}(\kappa)$, so the
probability of a key not being remapped to a different hash code is
$1/|C'|$, or $p(h_C(\kappa) = h_{C'}(\kappa)) = 1/|C'|$. Clearly, as
the set of hash codes grows large, the probability of a key mapping to
the same hash code after the addition or removal of a new hash code
approaches zero.

In many applications, this high likelihood of keys being remapped when
$C$ is altered is unacceptable; so we require a different hash
function. For example, in the case of a distributed cache, a hash
function might be being used to determine which node contains a
requested object, with the key provided to the hash function being an
object identifier (e.g. a URL), and the hash codes being node
identifiers (e.g. I.P. addresses). In this scenario, when the set of
nodes (i.e. hash codes) is altered, we want to minimise the number of
objects that must move between nodes in order to satisfy the new
mapping. Consistent Hashing \cite{Karger1997} provides exactly this:
in addition to the usual determinism and uniformity properties of hash
functions, it also requires that when a hash code is added or removed,
only the minimal number of keys are remapped to maintain
uniformity. In the case of addition, the minimal number of keys is
simply the number of keys that must be mapped to the new hash code;
therefore it is not permitted to remap keys between existing hash
codes.

This remapping requirement in combination with the uniformity
requirement reveals further details of how the hash function should
behave. For the uniformity property to be maintained after a hash code
is added, each existing hash code should give up an equal proportion
of their keys to become the keys of the new hash code. Once again with
$C \subset C'$ and $|C'| = |C| + 1$, the uniformity property gives us
$\forall \sigma \in C, \forall \kappa. \: p(h_C(\kappa) = \sigma) =
1/|C|$ and $\forall \sigma \in C', \forall \kappa. \: p(h_{C'}(\kappa)
= \sigma) = 1/|C'|$. We can now relate these by \[ \forall \sigma \in
C, \forall \kappa. \: h_C(\kappa) = \sigma \implies p(h_{C'}(\kappa) =
\sigma) = \frac{|C|}{|C'|} \] i.e. if $h_C(\kappa)$ yields hash code
$\sigma$ for key $\kappa$, then the probability of $h_{C'}(\kappa)$
yielding the same $\sigma$ for the same $\kappa$ is $|C| / |C'|$. The
probability of $h_{C'}(\kappa)$ yielding a hash code from $C$ is $|C|
/ |C'|$, leaving $1 / |C'|$ for the new hash code, as required. Note
that there is no possibility of keys being moved between existing hash
codes: each hash code loses only as many keys as are required to be
donated to the new hash code; all remaining keys stay with their
existing hash code thus the remapping property is satisfied. The
inverse specifies the conditions of a hash code being removed: to
maintain the uniformity property, the removed hash code's keys must be
equally distributed amongst the remaining hash codes, and to maintain
the remapping property, no keys are remapped except those that were
mapped to the now departed hash code. It is worth observing that the
requirement that only $1/|C'|$ of the keys are remapped is the
complement of that achieved by our unsuitable simple hash function in
which only $1/|C'|$ keys are {\em not} remapped!

The introduction of the properties of a Consistent Hash also presented
an algorithm \cite{Karger1997}, which we refer to as the classic
algorithm. Whilst this algorithm has been implemented in many
different programming languages and used in many scenarios, it has
several flaws which we explore in this paper. We then present a new
algorithm which precisely achieves the desired properties and solves
the flaws identified in the classic algorithm.

\section{Classic Algorithm}

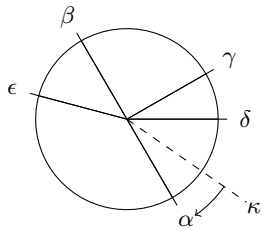
\begin{figure}
\begin{center}
\begin{tikzpicture}[scale=1.2]
\newcounter{a}
\newcounter{b}
\foreach \t/\l in
         {
           30/$\delta$,
           90/$\gamma$,
           45/$\beta$,
           135/$\epsilon$,
           60/$\alpha$
         }
         {
           \setcounter{a}{\value{b}}
           \addtocounter{b}{\t}
           \slice{\thea} {\theb} {\l}
         }

\draw[dashed] (0,0) -- (325:1.6);
\node at (325:1.7) {$\kappa$};
\draw[->] (0,0) (325:1.3) arc (325:305:1.3);
\end{tikzpicture}
\end{center}
\caption{A circle divided into five segments by five randomly placed
  hash code points, and the search for the {\em next} hash code point
  from the key $\kappa$}
\label{fig:circle-simple}
\end{figure}

The classic algorithm places hash codes at random points around a
circle. Keys to the hash function are interpreted as points on the
circle, and the hash function identifies the {\em next} hash code
point around the circle, \fref{fig:circle-simple}. In this way, we can
think of each hash code as owning different segments of the circle. It
is usual to apply a standard hash function to keys in order to limit
the keys to the range of the circle and ensure uniform distribution of
keys around the circle. Whilst the hash codes are placed at random
points around the circle, the points may still be found
deterministically, for example generated by applying a standard hash
function to the hash code names themselves.

Because each discrete point around the circle can only be occupied by
a single hash code, addition of hash codes is not commutative. This is
not an essential property, but it does have an effect on the size of
the state that must be maintained for the hash function. Addition is
not commutative because of the possibility that two hash codes both
try to occupy the same point around the circle. When such a collision
occurs, a number of solutions are possible, but one of the two hash
codes must {\em win} the particular point otherwise the remapping
property will be violated (commutativity could be achieved if neither
hash code wins the contested point, but that would result in keys
being transferred from the first hash code added (the initial winner),
to a different existing hash code, which is illegal). If hash code
points around the circle are randomly generated, then the state
maintained must include all those points (when a collision occurs, a
fresh random point is generated for the new hash code). If the hash
code points are generated deterministically by applying a standard
hash function to the hash code identifiers, then the state must
maintain the order in which hash codes were added and removed so that
the points can be correctly reestablished including the outcome of
collisions (when a collision occurs, a number of options are
available, such as hashing the concatenation of the new hash code name
and an attempt-number).

\begin{figure}
\begin{center}
\begin{align*}
\vcenter{
    \hbox{
      \begin{tikzpicture}[scale=1.2]
        \setcounter{b}{0}
        \foreach \t/\l in
                 {
                   30/$\delta$,
                   90/$\gamma$,
                   180/$\beta$,
                   60/$\alpha$
                 }
                 {
                   \setcounter{a}{\value{b}}
                   \addtocounter{b}{\t}
                   \segment{\thea} {\theb} {\l}
                 }
      \end{tikzpicture}
    }
}
&& \overset{+\epsilon}\Longrightarrow &&
\vcenter{
  \hbox{
    \begin{tikzpicture}[scale=1.2]
      \setcounter{b}{0}
      \foreach \t/\l in
               {
                 30/$\delta$,
                 90/$\gamma$,
                 45/$\beta$,
                 135/$\epsilon$,
                 60/$\alpha$
               }
               {
                 \setcounter{a}{\value{b}}
                 \addtocounter{b}{\t}
                 \segment{\thea} {\theb} {\l}
               }
    \end{tikzpicture}
  }
}
\end{align*}
\end{center}
\caption{Modifying an existing circle by the addition (left to right),
  or removal (right to left) of a point for the hash code $\epsilon$}
\label{fig:circle-simple-add}
\end{figure}
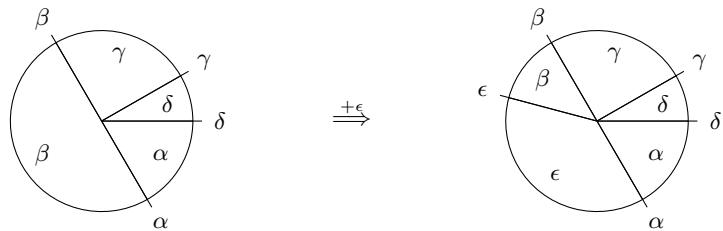

This classic algorithm achieves the remapping and deterministic
properties but fails to guarantee the uniformity property. Whenever a
hash code is added, its determined point around the circle splits an
existing segment, transferring a portion of the keys from that segment
to the new hash code. No other segments around the circle are altered
so all other keys remain mapped to their existing hash codes,
\fref{fig:circle-simple-add}. Similarly, when a hash code is removed,
the disappearance of its point around the circle merges its segment
with the {\em next} segment. Again, no other points around the circle
are altered, so no keys are remapped between other, surviving, hash
codes. As a result, the remapping property is achieved. The
determinism property is immediate, provided state is maintained
appropriately in light of changes to $C$ as discussed previously.

Uniformity however is at best approximated. The location of hash code
points is random, so with just two hash codes, it is very unlikely for
each hash code's points to be $180^\circ$ apart: just $1/R$ where $R$
is the range of the circle. So the likelihood of the segments being of
equal size and thus uniformity achieved, is very low. If one traverses
around the perimeter of the circle at constant speed, encountering a
hash code point (and entering a new segment) can be modelled by a
Poisson process. The interval between such encounters (i.e. the
segment length) is then an exponential distribution
\cite{Niedermayer2005,Xiaouien2009}. The extreme left-skew of this
distribution demonstrates how exceedingly unlikely it is to achieve
uniformity. For example, a circle divided by 10 points, one for each
of 10 hash codes, will have a mean arc length of $36^\circ$, but a
median arc length of just $25^\circ$.

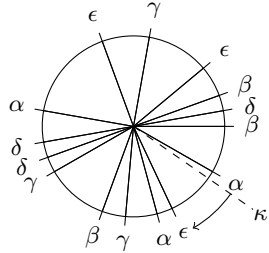
\begin{figure}
\begin{center}
\begin{tikzpicture}[scale=1.2]
\setcounter{b}{0}
\foreach \t/\l in
         {
           10/$\beta$,
           10/$\delta$,
           20/$\beta$,
           40/$\epsilon$,
           30/$\gamma$,
           60/$\epsilon$,
           20/$\alpha$,
           10/$\delta$,
           10/$\delta$,
           40/$\gamma$,
           15/$\beta$,
           20/$\gamma$,
           10/$\alpha$,
           35/$\epsilon$,
           30/$\alpha$
         }
         {
           \setcounter{a}{\value{b}}
           \addtocounter{b}{\t}
           \slice{\thea} {\theb} {\l}
         }

\draw[dashed] (0,0) -- (325:1.6);
\node at (325:1.7) {$\kappa$};
\draw[->] (0,0) (325:1.3) arc (325:300:1.3);
\end{tikzpicture}
\end{center}
\caption{A circle of segments for five hash codes, each with three
  random points ($k = 3$), being indexed by the key $\kappa$}
\label{fig:circle-multi-point}
\end{figure}

To address this, the classic algorithm uses several points for each
hash code, \fref{fig:circle-multi-point}. This changes the
distribution of segment lengths from exponential to Erlang; an
Erlang-$k$ distribution is the sum of $k$ independent exponential
distributions. If $k$ points are used per hash code then the
distribution of lengths forms an Erlang-$k$ distribution (the second
parameter to the Erlang distribution, $\lambda$, is in this case $k
\cdot |C|$). As $k$ rises, so the skew in the distribution of lengths
is reduced. However, the ratios of smallest and largest segment
lengths to the mean length remains high: with $k$ as high as $|C|$,
the mean length will be around $1.1$ of the smallest length, and the
largest length will also be around $1.1$ of the mean length. With $k =
4 \cdot |C|$, these ratios fall to around $1.06$, and with $k = 8
\cdot |C|$, they only fall to around $1.04$.

In some scenarios it may well be acceptable to have one hash code
receive $8\%$ more keys than another. Note though that this imbalance
is not reduced by adding additional hash codes, indeed quite the
opposite: the lower $k / |C|$ falls, the greater the spread. To reduce
the imbalance, the value of $k$ needs to be determined as a multiple
of the {\em maximum} number of hash codes. Thus if the set of hash
codes is normally relatively small and only under certain conditions
are many more hash codes added, then the circle will generally contain
many more points than strictly necessary, due to the high $k$. In
scenarios with large numbers of hash codes, the probability of points
colliding rises and you may even run out of points: with 20,000 hash
codes each with 200,000 points, over $93\%$ of 32-bit integers are
used up, thus switching to a 64-bit circle perimeter range would
become necessary.

As the classical algorithm is typically implemented by holding the
points in a binary tree, the depth of the tree determines the look-up
cost. Assuming any other hash functions being used to prepare the key
are $\BigO{1}$, we should have an overall cost of $\BigO{\log_2(k
  \cdot |C|)}$ which simplifies to $\BigO{\log_2(|C|)}$ as $k$ is a
constant. Some implementations claim worst cost of $\BigO{1}$ because
of the finite upper limit on the number of points around the circle,
and thus a limit on the depth of the corresponding tree, but that's
arguably a consequence of limitations of the implementation, and
average cost (rather than worst-case) is still
$\BigO{\log_2(|C|)}$. Techniques do exist to reduce this for the
average case, but they increase memory footprint and make adding and
removing hash codes more expensive. Obviously, this does not
invalidate such approaches, but we will not consider them further
here.

The memory footprint of so many points is also worth considering: if
an implementation holds the points in some sort of tree, 20,000 hash
codes each with 200,000 points would result in a tree of depth 32,
with 8.6 billion nodes (number of nodes in a tree is found by $2^{d+1}
- 1$ where $d$ is the tree depth). Assuming each node carries two
64-bit pointers to its children and a 64-bit value, we have a minimum
of 192 bits per node, or 24 bytes. Such a tree then works out at a
minimum memory cost of around 200 GB, or around 10 MB per hash code.

\section{Analysing Requirements}

The large number of replicas in the classic algorithm is not only
necessary to address the exponential distribution of segment lengths
(and thus approximate uniformity given a static number of hash codes)
but also to maintain approximate uniformity in light of changes to the
set of hash codes, $C$. We now consider how to algorithmically place
points around the circle for each hash code such that we precisely
achieve and maintain uniformity, and do not rely on any random
distributions.

As stated previously, when a hash code is added, it should inherit an
equal number of keys from each of the existing hash codes, and when it
leaves, it should equally distribute its keys to the surviving hash
codes. In the classic algorithm, to achieve this with the removal of a
hash code requires that that hash code must have at least as many
segments as there are remaining hash codes so that each of its own
segments might be followed by a segment of a each of the other hash
codes.\footnote{In the classic algorithm, due to the random placement
  of hash code points, a high number of replicas, $k$, is also
  necessary to have confidence the required permutations are achieved,
  but as we shall show, as the number of hash codes increases, the
  multiple of $|C|$ to define $k$ must itself rise.}

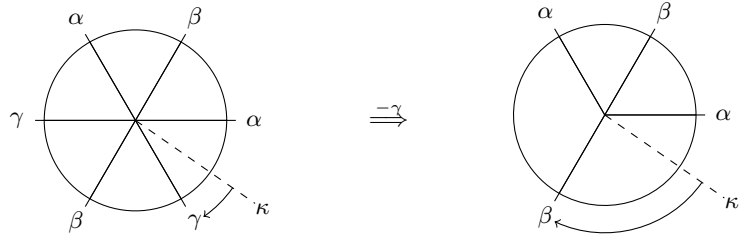
\begin{figure}
\begin{center}
\begin{align*}
\vcenter{
    \hbox{
      \begin{tikzpicture}[scale=1.2]
        \setcounter{b}{0}
        \foreach \t/\l in
                 {
                   60/$\alpha$,
                   60/$\beta$,
                   60/$\alpha$,
                   60/$\gamma$,
                   60/$\beta$,
                   60/$\gamma$
                 }
                 {
                   \setcounter{a}{\value{b}}
                   \addtocounter{b}{\t}
                   \slice{\thea} {\theb} {\l}
                 }
                 \draw[dashed] (0,0) -- (325:1.6);
                 \node at (325:1.7) {$\kappa$};
                 \draw[->] (0,0) (325:1.3) arc (325:305:1.3);
      \end{tikzpicture}
    }
}
&& \overset{-\gamma}\Longrightarrow &&
\vcenter{
  \hbox{
      \begin{tikzpicture}[scale=1.2]
        \setcounter{b}{0}
        \foreach \t/\l in
                 {
                   60/$\alpha$,
                   60/$\beta$,
                   120/$\alpha$,
                   120/$\beta$
                 }
                 {
                   \setcounter{a}{\value{b}}
                   \addtocounter{b}{\t}
                   \slice{\thea} {\theb} {\l}
                 }
                 \draw[dashed] (0,0) -- (325:1.6);
                 \node at (325:1.7) {$\kappa$};
                 \draw[->] (0,0) (325:1.3) arc (325:245:1.3);
      \end{tikzpicture}
  }
}
\end{align*}
\end{center}
\caption{A circle containing all possible pairs of three hash codes
  ($\alpha$, $\beta$, $\gamma$) as neighbouring segments of equal size
  (left), being modified by the removal of a hash code $\gamma$
  (right), both being indexed by the key $\kappa$}
\label{fig:circle-perfect-3}
\end{figure}

For example, with three hash codes, $\alpha$, $\beta$, $\gamma$, we
require that $\alpha$ must have at least two segments: one followed by
$\beta$ and the other followed by $\gamma$. The same holds for the
other two hash codes, so we must accommodate all possible pairs around
the circle: $(\alpha, \beta)$, $(\beta, \alpha)$, $(\alpha, \gamma)$,
$(\gamma, \alpha)$, $(\beta, \gamma)$, $(\gamma, \beta)$. One solution
would be $[\alpha, \beta, \alpha, \gamma, \beta, \gamma]$, with each
segment being of equal length (\fref{fig:circle-perfect-3}), though
there are a number of equivalent solutions. Note how the last element
of each pair forms the first element of a different pair, and thus
there are six segments of the circle corresponding to the six
pairs. This is a minimal solution: the required pairings cannot be
achieved with fewer points around the circle. If two hash codes leave,
by elimination, the one remaining hash code will inherit all the keys
and therefore we do not need to worry about the distribution of a
single key's points around the circle. This explains why with three
hash codes, we concern ourselves with pairs of hash codes, and not
triples.

With four hash codes, $\alpha$, $\beta$, $\gamma$, $\delta$, if two
hash codes leave, we still require uniformity of distribution of keys
to the two remaining hash codes. This is no longer about incorporating
every possible {\em pair} of hash codes around the circle, it is now
about incorporating every possible {\em triple} of hash codes: pairs
will maintain uniformity in case of one removal, triples are required
for two removals. In general, for $|C|$ hash codes, every permutation
of length $|C| - 1$ must exist around the circle in such a way that
all but the first element (i.e. the last $|C| - 2$ elements) of each
permutation forms the first elements of the next permutation. This is
known as a {\em universal cycle} for the $|C| - 1$ permutations of
$C$, and is a well studied problem \cite{Jackson1993}.

Permutations of length one less than the number of available symbols
are known as {\em shorthand} permutations as the remaining symbol is
implicit. For example with the symbols $\alpha$, $\beta$, $\gamma$,
$\delta$, the permutation $[\gamma,\delta,\beta]$ can be considered
shorthand for $[\gamma,\delta,\beta,\alpha]$. Thus we can also say
that we require a universal cycle of the shorthand permutations of
$C$. In general, it is always possible to find a universal cycle of
shorthand permutations, and efficient algorithms exist to directly
construct such permutations \cite{Ruskey,Holroyd2010}.

Such a cycle will work as desired in light of multiple removals of
hash codes. As every hash code in the cycle is followed an equal
number of times by each of the other hash codes, removal of any hash
code will equally distribute its keys amongst the remaining hash
codes, who's segments will grow in size. For example, with four hash
codes, $\alpha$, $\beta$, $\gamma$, $\delta$, a universal cycle of
shorthand permutations is:

\[ [\alpha,\beta,\gamma,\alpha,\beta,\delta,\alpha,\gamma,\beta,\alpha,\gamma,\delta,\beta,\alpha,\delta,\gamma,\beta,\delta,\gamma,\alpha,\delta,\beta,\gamma,\delta] \]

\begin{figure}
\begin{center}
\begin{tikzpicture}[scale=1.2]
\setcounter{b}{0}
\foreach \t/\l in
         {
           15/$\alpha$,
           15/$\beta$,
           15/$\gamma$,
           15/$\alpha$,
           15/$\beta$,
           15/$\delta$,
           15/$\alpha$,
           15/$\gamma$,
           15/$\beta$,
           15/$\alpha$,
           15/$\gamma$,
           15/$\delta$,
           15/$\beta$,
           15/$\alpha$,
           15/$\delta$,
           15/$\gamma$,
           15/$\beta$,
           15/$\delta$,
           15/$\gamma$,
           15/$\alpha$,
           15/$\delta$,
           15/$\beta$,
           15/$\gamma$,
           15/$\delta$
         }
         {
           \setcounter{a}{\value{b}}
           \addtocounter{b}{\t}
           \slice{\thea} {\theb} {\l}
         }
\end{tikzpicture}
\end{center}
\caption{A circle constructed from the 24 elements of a universal
  cycle of the shorthand permutations of four hash codes, $\alpha$,
  $\beta$, $\gamma$ and $\delta$}
\label{fig:circle-perfect-4}
\end{figure}
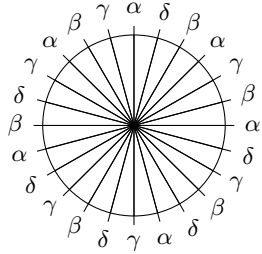

\noindent
Uniformity is achieved: each hash code has six entries and thus six
segments of equal length around the circle,
\fref{fig:circle-perfect-4}. If we remove the hash code $\gamma$ then
we are left with:

\[ [\alpha,\beta,\alpha,\alpha,\beta,\delta,\alpha,\beta,\beta,\alpha,\delta,\delta,\beta,\alpha,\delta,\beta,\beta,\delta,\alpha,\alpha,\delta,\beta,\delta,\delta] \]

\begin{figure}
\begin{center}
\begin{align*}
\begin{tikzpicture}[scale=1.2]
\setcounter{b}{0}
\foreach \t/\l in
         {
           15/$\alpha$,
           30/$\beta$,
           15/$\alpha$,
           15/$\beta$,
           15/$\delta$,
           30/$\alpha$,
           15/$\beta$,
           30/$\alpha$,
           15/$\delta$,
           15/$\beta$,
           15/$\alpha$,
           30/$\delta$,
           15/$\beta$,
           30/$\delta$,
           15/$\alpha$,
           15/$\delta$,
           30/$\beta$,
           15/$\delta$
         }
         {
           \setcounter{a}{\value{b}}
           \addtocounter{b}{\t}
           \slice{\thea} {\theb} {\l}
         }
\end{tikzpicture}
&&
\begin{tikzpicture}[scale=1.2]
\setcounter{b}{0}
\foreach \t/\l in
         {
           15/$\alpha$,
           30/$\beta$,
           15/$\alpha$,
           30/$\beta$,
           30/$\alpha$,
           15/$\beta$,
           45/$\alpha$,
           15/$\beta$,
           45/$\alpha$,
           45/$\beta$,
           30/$\alpha$,
           45/$\beta$
         }
         {
           \setcounter{a}{\value{b}}
           \addtocounter{b}{\t}
           \slice{\thea} {\theb} {\l}
         }
\end{tikzpicture}
\end{align*}
\end{center}
\caption{The circle of \fref{fig:circle-perfect-4} with the hash codes
  $\gamma$ removed (left), and $\gamma$ and $\delta$ removed (right)}
\label{fig:circle-perfect-4-remove}
\end{figure}
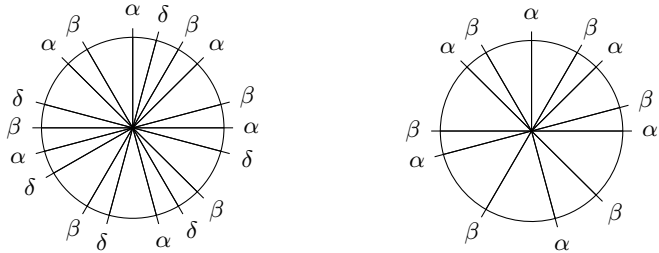

\noindent
In our notation here, removal is represented by substitution of the
removed hash code with the next surviving hash code. As each element
is a segment of the circle of equal length, this makes it easier to
check uniformity; each remaining hash code now has eight entries, and
thus uniformity has been maintained
(\fref{fig:circle-perfect-4-remove}).

Note how there are two $\alpha,\alpha$ pairs, two $\beta,\beta$ pairs,
and two $\delta,\delta$ created by the removal of the six $\gamma$
segments. Each of these themselves are still followed by each of the
remaining hash codes: one $\alpha,\alpha$ pair is followed by a
$\beta$, and one by a $\delta$; similarly for the other pairs. Thus
multiple removals of hash codes still result in uniformity of
distribution of keys. This is not a surprising result given the nature
of the permutations: the very reason why there are two instances of
$\gamma$ followed by $\alpha$ (thus forming the two $\alpha,\alpha$
pairs upon removal of $\gamma$) is so that they can be followed by
each of the remaining hash codes: $\beta$ and $\delta$ in this case.

If we remove another hash code, for example $\delta$, then we are now
left with:

\[ [\alpha,\beta,\alpha,\alpha,\beta,\alpha,\alpha,\beta,\beta,\alpha,\beta,\beta,\beta,\alpha,\beta,\beta,\beta,\alpha,\alpha,\alpha,\beta,\beta,\alpha,\alpha] \]

\noindent
Again, uniformity has been maintained: we have now 12 $\alpha$s and 12
$\beta$s (\fref{fig:circle-perfect-4-remove}).

Whilst it is clear that removal will maintain uniformity when the
circle is constructed from a universal cycle of shorthand permutations
of $C$, it is less clear how to construct such circles additively:
given the cycle for $\alpha$, $\beta$, $\gamma$, $\delta$ given in
\fref{fig:circle-perfect-4}, how do you modify it to incorporate a new
hash code, $\epsilon$, whilst achieving the remapping property?
Equivalently, given the circle diagram of the remaining $\alpha$s and
$\beta$s on the right in \fref{fig:circle-perfect-4-remove}, it is far
from clear how to construct this, and thus create the necessary spaces
for later hash codes to fill. In the classic algorithm, the circle
exists to achieve the remapping property both for addition and removal
of hash codes. However, with the positions of each hash code point
being precisely determined by the universal cycle, the circle only now
serves to provide the remapping property upon removal of a hash code,
not the addition: addition can no longer be achieved by splitting
existing segments. Happily, our new algorithm manages to achieve the
uniformity and remapping properties precisely, without needing to
address this problem.

However, first let us examine the size of these cycles. Because all
but one hash code from each permutation overlaps with the next hash
code, each permutation contributes one hash code to the length of the
cycle. The number of shorthand permutations of $C$ is the same as the
number of $|C|$-length permutations of $C$, which is $|C|!$. Thus with
just 12 hash codes, we have a cycle length of 479,001,600. Whilst this
is less than $2^{32}$, 13 hash codes would be create a cycle length
greater than $2^{32}$. This not only impacts the representation of the
circle (and its memory footprint if the entire circle must be
constructed and maintained), but also affects the key: in essence what
this means is that a 32-bit key can only choose between up to 12 hash
codes. A 512-bit key can only choose between up to 98 hash codes. This
has implications for consistent hashing generally: with large numbers
of hash codes and short keys, it is impossible to achieve perfect
uniformity, and an approximate solution in such scenarios cannot be
bettered. To achieve perfect uniformity and the remapping property in
light of removals, every permutation needs an equal chance of being
selected. The use of shorthand permutations is only necessary to be
able to construct universal cycles out of the segments around the
circle.

This factorial of $|C|$ also impacts performance. As discussed
earlier, the classic algorithm is typically implemented using a binary
tree to hold the points around the circle. The depth of the tree and
thus the average cost of look-up is now $\BigO{\log_2(|C|!)}$ which is
worse than $\BigO{|C|}$ (we present an intuitive proof of this
later). However, with even small numbers of hash codes, the factorial
results in so many nodes that it is unwise to maintain the whole tree
in memory. Instead the nodes of the tree would need to be constructed
by some means as the tree was traversed. This would likely result in a
very different look-up cost.

The factorial also explains why the multiple of $|C|$ to define the
number of replicas, $k$, in the classic algorithm must rise itself as
$|C|$ rises: to approximate maintaining uniformity in light of
removals, the classic algorithm must have sufficient points per hash
code to approximate a universal cycle of shorthand permutations of
$C$. Thus $k$ should also be a multiple of the factorial of $|C|$ to
have confidence of being able to approximate such a cycle by random
placement of hash code points.

\section{New algorithm}

In the previous section, the universal cycle served to position the
hash codes around the circle such that uniformity was achieved, and
that in the event of removal of hash codes, uniformity would be
maintained. It can be considered that what the hash function is
actually returning is not a single hash code, but a permutation of the
hash codes, with removed hash codes filtered out.

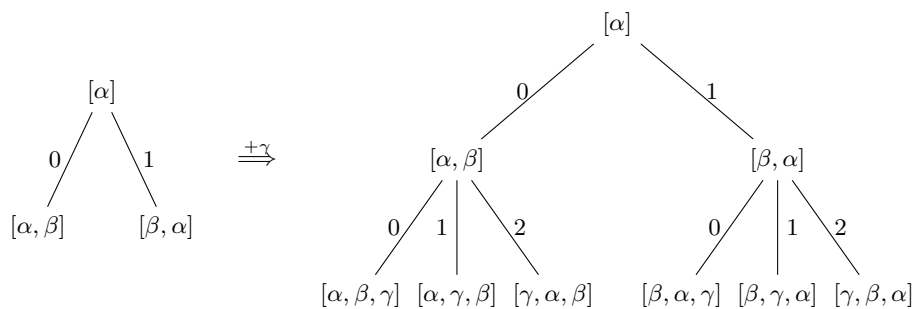
\begin{figure}
\begin{center}
\begin{align*}
\vcenter{
    \hbox{
      \begin{tikzpicture}[scale=1.2,
          level 1/.style={sibling distance=40}]
        \node {$[\alpha]$}
          child {node {$[\alpha,\beta]$} edge from parent node[left] {0}}
          child {node {$[\beta,\alpha]$} edge from parent node[right] {1}};
      \end{tikzpicture}
    }
}
&& \overset{+\gamma}\Longrightarrow &&
\vcenter{
  \hbox{
      \begin{tikzpicture}[scale=1.2,
          level 1/.style={sibling distance=100},
          level 2/.style={sibling distance=30}]
        \node {$[\alpha]$}
          child {node {$[\alpha,\beta]$}
            child {node {$[\alpha,\beta,\gamma]$} edge from parent node[left] {0}}
            child {node {$[\alpha,\gamma,\beta]$} edge from parent node[left] {1}}
            child {node {$[\gamma,\alpha,\beta]$} edge from parent node[right] {2}}
            edge from parent node[left] {0}
          }
          child {node {$[\beta,\alpha]$}
            child {node {$[\beta,\alpha,\gamma]$} edge from parent node[left] {0}}
            child {node {$[\beta,\gamma,\alpha]$} edge from parent node[right] {1}}
            child {node {$[\gamma,\beta,\alpha]$} edge from parent node[right] {2}}
            edge from parent node[right] {1}
          };
      \end{tikzpicture}
  }
}
\end{align*}
\end{center}
\caption{A tree of the permutations of two hash codes (left), being extended by an additional layer for a third hash code (right)}
\label{fig:tree-2-add}
\end{figure}

Our new algorithm explicitly returns a permutation of all the hash
codes. The interpretation of a permutation as a result of the hash
function is not fixed, but for our purposes, we read $[\alpha,\beta]$
as {\em first try $\alpha$, then try $\beta$}. Every permutation is
equally probable, which achieves both the uniformity requirement and
the remapping requirement in light of removal of hash codes (i.e. as
before, filtering out removed hash codes from the resulting
permutation will maintain uniformity). Each permutation exists as a
leaf of a tree, but this is not a binary tree: whilst the root node
has two children, all other nodes have one more child than does their
parent, \fref{fig:tree-2-add}. We then subdivide the key to navigate
through the tree. The remapping requirement means that if we use part
of the key to decide between different orderings of particular hash
codes in the resulting permutation then we must forevermore use that
same part of the key to make that same decision.

With one hash code, the result is trivial. With two hash codes,
$\alpha$ and $\beta$, we want the answer to be the permutation
$[\alpha,\beta]$ as often as the permutation $[\beta,\alpha]$. To
choose between these, we use the key ($\kappa$) modulus two. With
three hash codes, $\alpha$, $\beta$ and $\gamma$, we now have six
permutations. As we previously used $\kappa \bmod 2$ to choose between
{\em $\alpha$ before $\beta$} versus {\em $\beta$ before $\alpha$}, we
must continue to do so, and must then discard that part of the key (by
dividing by two). At the next layer of the tree we have two 3-way
choices, each refining the previous choice by adding in the new hash
code $\gamma$. Thus we use the remaining key modulus three to make
this choice. The tree at the right of \fref{fig:tree-2-add} is shown
as a table in \fref{fig:tree-3-table}.

\begin{figure}
\begin{center}
\begin{tabular}{|c|c|c|c|}
\hline $\kappa \bmod 6$ & $\frac{\kappa}{2} \bmod 3$ & $\kappa \bmod 2$ & Permutation \\
\hline 0 & 0 & 0 & $[\alpha,\beta,\gamma]$ \\
\hline 1 & 0 & 1 & $[\beta,\alpha,\gamma]$ \\
\hline 2 & 1 & 0 & $[\alpha,\gamma,\beta]$ \\
\hline 3 & 1 & 1 & $[\beta,\gamma,\alpha]$ \\
\hline 4 & 2 & 0 & $[\gamma,\alpha,\beta]$ \\
\hline 5 & 2 & 1 & $[\gamma,\beta,\alpha]$ \\\hline
\end{tabular}
\end{center}
\caption{The tree on the right of \fref{fig:tree-2-add} as a table for ease of reading}
\label{fig:tree-3-table}
\end{figure}

In general, each layer of the tree adds a new hash code. A node of any
particular layer can be seen to receive a permutation from its parent,
and to add its new hash code in every possible position within that
permutation; each node has a child for each of the possible positions
at which its own hash code can be inserted. But no modifications are
made to the ordering of existing hash codes within the permutation
which a node receives, and it is this that achieves the remapping
property. Note that in the tree of \fref{fig:tree-2-add}, the index of
each branch indicates the distance from the {\em end} of the existing
permutation at which the new hash code is inserted. This is an
arbitrary choice: any strategy for determining the position of the new
hash code within the received permutation is acceptable (and can even
vary per layer), provided the strategy is both deterministic and
uniform.

If we consider all keys modulus two (i.e. the values $0$ and $1$),
then with just the hash codes $\alpha$ and $\beta$, we see $0$ (and
thus all even keys) maps to $[\alpha,\beta]$, and $1$ (and thus all
odd keys) maps to $[\beta,\alpha]$. With the hash codes $\alpha$,
$\beta$ and $\gamma$, all keys modulus six (i.e. the values $0$ to
$5$), and if we just consider the first element of each permutation
returned, then we see $0$ and $2$ are mapped to $\alpha$ (as they were
previously without the $\gamma$ hash code), $1$ and $3$ are mapped to
$\beta$ (as they were previously without the $\gamma$ hash code), and
$4$ and $5$ are mapped to $\gamma$. Thus we have ensured that in the
transition from two to three hash codes, we only remap keys to the new
value ($4$ and $5$ going to $\gamma$), and we have taken an equal (and
minimal) number of keys from each of the existing hash codes (i.e. $4$
from $\alpha$ and $5$ from $\beta$), resulting in an equal
distribution of keys to values. Uniformity has been maintained and the
remapping property achieved.

We must also check what happens when a hash code is removed. If the
hash code removed is the most recently added, then we can simply
discard the lowest layer of the tree and return to the earlier
configuration. Otherwise, we continue with the existing tree, but must
filter out the removed hash code from the resulting permutation
(depending on the implementation, this filtering could be done as the
resulting permutation is constructed). With the hash codes $\alpha$,
$\beta$ and $\gamma$, there are two permutations which start with
$\alpha$: $[\alpha,\beta,\gamma]$ and $[\alpha,\gamma,\beta]$. If the
$\alpha$ hash code is removed, we see that its keys are equally
redistributed and uniformity maintained: we have one permutation where
the initial $\alpha$ is followed by a $\beta$ and one where it is
followed by a $\gamma$. As in the previous section, this is simply a
consequence of using permutations. It should be noted that in common
with the classic algorithm, our algorithm makes addition of hash codes
non-commutative: each hash code is accommodated by individual layers
of the tree (and thus is navigated by specific sections of the key),
and so the order in which the hash codes were added matters.

If after $\alpha$ has been removed we add the hash code $\delta$, then
$\delta$ can simply take the space created by the removal of
$\alpha$. The remapping and uniformity properties are precisely
achieved. We now can define exactly the state that our algorithm
requires: a list containing current hash codes interspersed with a
marker used to indicate free slots caused by hash codes being
removed. The list will only grow when a hash code is added and there
are no free slots, and will shrink whenever the hash code at the end
of the list is removed (thus you should never have a free slot marker
at the end of the list). Barring substitutions of a new hash code for
a free slot marker, the list elements will be in the order in which
the hash codes were added, corresponding to the layers of the tree.

The average cost of look-up is now $\BigO{|C|}$, as each hash code
adds a layer to the tree. Whilst we have the same number of leaves as
in the previous section (i.e. $|C|!$), we no longer use a binary tree:
each node has one more child than does its parent. As a child is
determined by indexing a node's list of children by the modulus of
part of the key, the selection of the child remains $\BigO{1}$ despite
nodes having increasing numbers of children as depth
increases. Consequently, we have one division, one modulus, and one
indexing operation per layer of the tree. Assuming each of these are
$\BigO{1}$, the average cost of reaching a leaf, and thus a look-up,
is $\BigO{|C|}$. This then is our intuitive proof that
$\BigO{\log_2(|C|!)}$ is worse than $\BigO{|C|}$: the binary tree from
the previous section and our non-binary tree from this section both
contain the same number of leaves, but the binary tree is limited to
two children per node and so must use more nodes than our non-binary
tree which has an additional child per node per
generation. Consequently, for the same number of leaves, the binary
tree must be deeper than our non-binary tree, thus the cost of
navigating to a leaf must be higher. Therefore $\BigO{\log_2(|C|!)}$
is worse than $\BigO{|C|}$.

As we have the same number of leaf nodes as in the previous section,
caused by the factorial of $|C|$, we have the same implications in the
relationship between key bit length (or entropy) and the permissible
number of hash codes. Thus whilst our algorithm does indeed achieve
perfect uniformity and satisfies the remapping property, the cost is
in the factorial relationship between the number of hash codes, and
the range of the key. Equally, the number of nodes in our tree is
given by $\sum\nolimits_{i=1}^{|C|}{i!}$. Whilst this is fewer nodes
than the binary tree for the same number of leaves (indeed, the number
of nodes of our tree tends towards half the number of nodes of the
binary tree), nevertheless the number of nodes makes it impractical to
maintain such a tree in memory, so once again we must construct the
permutation dynamically as we descend the tree. Such an implementation
is given in \fref{fig:code:new}. The performance cost will change
however: each layer of the tree will insert its hash code into the
resulting permutation. The most efficient mechanism for doing this
will be to build the permutation in a tree, for which insertions will
on average $\BigO{\log_2(n)}$ where $n$ is the number of values in the
tree. As we know we will need to do $|C|$ insertions and the average
number of values in the tree will be $|C|/2$, we have a total cost of
all the insertions of $\BigO{|C| \cdot \log_2(|C|/2)}$. Whilst this is
a worse average cost than navigating a pre-constructed tree (which was
$\BigO{|C|}$), the memory savings are significant.

\begin{figure}
\noindent
\ttfamily
\hlkwd{consistent\textunderscore hash}\hlstd{}\hlopt{(}\hlstd{HashCodes}\hlopt{,\ }\hlstd{Key}\hlopt{)\ {-}$>$}\hspace*{\fill}\\
\hlstd{}\hlstd{\ \ \ \ }\hlstd{}\hlkwd{consistent\textunderscore hash}\hlstd{}\hlopt{({[}{]},\ }\hlstd{Key}\hlopt{,\ }\hlstd{HashCodes}\hlopt{,\ }\hlstd{}\hlnum{1}\hlstd{}\hlopt{).}\hspace*{\fill}\\
\hlstd{}\hspace*{\fill}\\
\hlkwd{consistent\textunderscore hash}\hlstd{}\hlopt{(}\hlstd{Permutation}\hlopt{,\ }\hlstd{\textunderscore Key}\hlopt{,\ {[}{]},\ }\hlstd{\textunderscore CurrentBase}\hlopt{)\ {-}$>$}\hspace*{\fill}\\
\hlstd{}\hlstd{\ \ \ \ }\hlstd{Permutation}\hlopt{;}\hspace*{\fill}\\
\hlstd{}\hlkwd{consistent\textunderscore hash}\hlstd{}\hlopt{(}\hlstd{Permutation}\hlopt{,\ }\hlstd{Key}\hlopt{,\ {[}}\hlstd{HC\ \textbar \ HashCodes}\hlopt{{]},\ }\hlstd{CurrentBase}\hlopt{)\ {-}$>$}\hspace*{\fill}\\
\hlstd{}\hlstd{\ \ \ \ }\hlstd{Pos\ }\hlopt{=\ }\hlstd{Key\ rem\ CurrentBase}\hlopt{,}\hspace*{\fill}\\
\hlstd{}\hlstd{\ \ \ \ }\hlstd{Permutation1\ }\hlopt{=\ }\hlstd{}\hlkwd{insert\textunderscore at}\hlstd{}\hlopt{(}\hlstd{Permutation}\hlopt{,\ }\hlstd{HC}\hlopt{,\ }\hlstd{Pos}\hlopt{),}\hspace*{\fill}\\
\hlstd{}\hlstd{\ \ \ \ }\hlstd{}\hlkwd{consistent\textunderscore hash}\hlstd{}\hlopt{(}\hlstd{Permutation1}\hlopt{,\ }\hlstd{Key\ div\ CurrentBase}\hlopt{,\ }\hlstd{HashCodes}\hlopt{,}\hspace*{\fill}\\
\hlstd{}\hlstd{\ \ \ \ \ \ \ \ \ \ \ \ \ \ \ \ \ \ \ \ }\hlstd{CurrentBase\ }\hlopt{+\ }\hlstd{}\hlnum{1}\hlstd{}\hlopt{).}\hspace*{\fill}\\
\hlstd{}\hspace*{\fill}\\
\hlkwd{insert\textunderscore at}\hlstd{}\hlopt{(}\hlstd{List}\hlopt{,\ }\hlstd{E}\hlopt{,\ }\hlstd{Nth}\hlopt{)\ {-}$>$}\hspace*{\fill}\\
\hlstd{}\hlstd{\ \ \ \ }\hlstd{}\hlkwd{insert\textunderscore at}\hlstd{}\hlopt{(}\hlstd{E}\hlopt{,\ }\hlstd{Nth}\hlopt{,\ {[}{]},\ }\hlstd{List}\hlopt{).}\hspace*{\fill}\\
\hlstd{}\hspace*{\fill}\\
\hlkwd{insert\textunderscore at}\hlstd{}\hlopt{(}\hlstd{E}\hlopt{,\ }\hlstd{}\hlnum{0}\hlstd{}\hlopt{,\ }\hlstd{HeadRev}\hlopt{,\ }\hlstd{Tail}\hlopt{)\ {-}$>$}\hspace*{\fill}\\
\hlstd{}\hlstd{\ \ \ \ }\hlstd{}\hlkwd{lists}\hlstd{}\hlopt{:}\hlstd{}\hlkwd{reverse}\hlstd{}\hlopt{(}\hlstd{HeadRev}\hlopt{,\ {[}}\hlstd{E\ \textbar \ Tail}\hlopt{{]});}\hspace*{\fill}\\
\hlstd{}\hlkwd{insert\textunderscore at}\hlstd{}\hlopt{(}\hlstd{E}\hlopt{,\ }\hlstd{N}\hlopt{,\ }\hlstd{HeadRev}\hlopt{,\ {[}}\hlstd{Elem\ \textbar \ Tail}\hlopt{{]})\ {-}$>$}\hspace*{\fill}\\
\hlstd{}\hlstd{\ \ \ \ }\hlstd{}\hlkwd{insert\textunderscore at}\hlstd{}\hlopt{(}\hlstd{E}\hlopt{,\ }\hlstd{N\ }\hlopt{{-}\ }\hlstd{}\hlnum{1}\hlstd{}\hlopt{,\ {[}}\hlstd{Elem\ \textbar \ HeadRev}\hlopt{{]},\ }\hlstd{Tail}\hlopt{).}\hspace*{\fill}\\
\mbox{}
\normalfont
\normalsize
\caption{An implementation in Erlang of our new algorithm}
\label{fig:code:new}
\end{figure}

In the code listing of \fref{fig:code:new}, as we subdivide the key we
build up the permutation in a list rather than a tree. Whilst this
will be less efficient than using a tree, for small values of $|C|$
the difference will be slight and the code simplified (as ever, beware
large constant overheads!). Note that there is no marker provided to
indicate removed hash codes; instead these can be filtered out from
the resulting permutation as necessary. In this implementation, the
position calculated by each layer of the tree is the distance from the
{\em start} of the received permutation at which to insert the new
hash code. Thus this will produce permutations in a different order to
that of \fref{fig:tree-3-table} but the properties still hold, and the
code is simplified.

If there is no need to return a permutation, and instead only the
first element of the permutation is required as a result then further
simplifications can be made to the code by avoiding construction of
the permutation, and so reducing the average complexity back to
$\BigO{|C|}$. We keep track of the {\em current} first element of the
permutation, and update it at each layer if we find the position of
new hash code is 0, thus replacing the old first element. This is
shown in \fref{fig:code:new-one-result}.

\begin{figure}
\noindent
\ttfamily
\hlkwd{consistent\textunderscore hash}\hlstd{}\hlopt{({[}}\hlstd{HC\ \textbar \ HashCodes}\hlopt{{]},\ }\hlstd{Key}\hlopt{)\ {-}$>$}\hspace*{\fill}\\
\hlstd{}\hlstd{\ \ \ \ }\hlstd{}\hlkwd{consistent\textunderscore hash}\hlstd{}\hlopt{(}\hlstd{HC}\hlopt{,\ }\hlstd{Key}\hlopt{,\ }\hlstd{HashCodes}\hlopt{,\ }\hlstd{}\hlnum{2}\hlstd{}\hlopt{).}\hspace*{\fill}\\
\hlstd{}\hspace*{\fill}\\
\hlkwd{consistent\textunderscore hash}\hlstd{}\hlopt{(}\hlstd{Result}\hlopt{,\ }\hlstd{\textunderscore Key}\hlopt{,\ {[}{]},\ }\hlstd{\textunderscore CurrentBase}\hlopt{)\ {-}$>$}\hspace*{\fill}\\
\hlstd{}\hlstd{\ \ \ \ }\hlstd{Result}\hlopt{;}\hspace*{\fill}\\
\hlstd{}\hlkwd{consistent\textunderscore hash}\hlstd{}\hlopt{(}\hlstd{Result}\hlopt{,\ }\hlstd{Key}\hlopt{,\ {[}}\hlstd{HC\ \textbar \ HashCodes}\hlopt{{]},\ }\hlstd{CurrentBase}\hlopt{)\ {-}$>$}\hspace*{\fill}\\
\hlstd{}\hlstd{\ \ \ \ }\hlstd{Result1\ }\hlopt{=\ }\hlstd{}\hlkwa{case\ }\hlstd{Key\ rem\ CurrentBase\ }\hlkwa{of}\hspace*{\fill}\\
\hlstd{}\hlstd{\ \ \ \ \ \ \ \ \ \ \ \ \ \ \ \ \ \ }\hlstd{}\hlnum{0\ }\hlstd{}\hlopt{{-}$>$\ }\hlstd{HC}\hlopt{;}\hspace*{\fill}\\
\hlstd{}\hlstd{\ \ \ \ \ \ \ \ \ \ \ \ \ \ \ \ \ \ }\hlstd{\textunderscore \ }\hlopt{{-}$>$\ }\hlstd{Result\hspace*{\fill}\\
}\hlstd{\ \ \ \ \ \ \ \ \ \ \ \ \ \ }\hlstd{}\hlkwa{end}\hlstd{}\hlopt{,}\hspace*{\fill}\\
\hlstd{}\hlstd{\ \ \ \ }\hlstd{}\hlkwd{consistent\textunderscore hash}\hlstd{}\hlopt{(}\hlstd{Result1}\hlopt{,\ }\hlstd{Key\ div\ CurrentBase}\hlopt{,\ }\hlstd{HashCodes}\hlopt{,}\hspace*{\fill}\\
\hlstd{}\hlstd{\ \ \ \ \ \ \ \ \ \ \ \ \ \ \ \ \ \ \ \ }\hlstd{CurrentBase\ }\hlopt{+\ }\hlstd{}\hlnum{1}\hlstd{}\hlopt{).}\hspace*{\fill}\\
\mbox{}
\normalfont
\normalsize
\caption{Simplifications achieved by only requiring a single result}
\label{fig:code:new-one-result}
\end{figure}

However, in this simpler scenario, we need to be much more careful
about removed hash codes: we cannot permit the algorithm to return the
marker for a removed hash code as the result, as we have no way of
knowing what would have been next in the permutation. Instead, we must
cope with the removed-hash-code marker directly in the implementation
itself. This is trickier as we need to consider many different
combinations of each layer updating the current result.
\Fref{fig:code:new-one-result-marker} shows an example
implementation. In the input list of hash codes, the Erlang atom
{\ttfamily\hlstd{undefined}\normalfont} is used to indicate the
removed-hash-code marker.

\begin{figure}
\noindent
\ttfamily
\hlkwd{consistent\textunderscore hash}\hlstd{}\hlopt{(}\hlstd{HashCodes}\hlopt{,\ }\hlstd{Key}\hlopt{)\ {-}$>$}\hspace*{\fill}\\
\hlstd{}\hlstd{\ \ \ \ }\hlstd{}\hlkwd{consistent\textunderscore hash}\hlstd{}\hlopt{(}\hlstd{undefined}\hlopt{,\ }\hlstd{Key}\hlopt{,\ }\hlstd{HashCodes}\hlopt{,\ }\hlstd{}\hlnum{1}\hlstd{}\hlopt{).}\hspace*{\fill}\\
\hlstd{}\hspace*{\fill}\\
\hlkwd{consistent\textunderscore hash}\hlstd{}\hlopt{(\{}\hlstd{HC}\hlopt{,\ }\hlstd{\textunderscore Pos}\hlopt{\},\ }\hlstd{\textunderscore Key}\hlopt{,\ {[}{]},\ }\hlstd{\textunderscore CurrentBase}\hlopt{)\ {-}$>$}\hspace*{\fill}\\
\hlstd{}\hlstd{\ \ \ \ }\hlstd{HC}\hlopt{;}\hspace*{\fill}\\
\hlstd{}\hlkwd{consistent\textunderscore hash}\hlstd{}\hlopt{(\{}\hlstd{candidate}\hlopt{,\ }\hlstd{HC}\hlopt{,\ }\hlstd{\textunderscore PosC}\hlopt{,\ }\hlstd{\textunderscore PosI}\hlopt{\},\ }\hlstd{\textunderscore Key}\hlopt{,\ {[}{]},\ }\hlstd{\textunderscore CurrentBase}\hlopt{)\ {-}$>$}\hspace*{\fill}\\
\hlstd{}\hlstd{\ \ \ \ }\hlstd{HC}\hlopt{;}\hspace*{\fill}\\
\hlstd{}\hlkwd{consistent\textunderscore hash}\hlstd{}\hlopt{(}\hlstd{Result}\hlopt{,\ }\hlstd{Key}\hlopt{,\ {[}}\hlstd{HC\ \textbar \ HashCodes}\hlopt{{]},\ }\hlstd{CurrentBase}\hlopt{)\ {-}$>$}\hspace*{\fill}\\
\hlstd{}\hlstd{\ \ \ \ }\hlstd{PosN\ }\hlopt{=\ }\hlstd{Key\ rem\ CurrentBase}\hlopt{,}\hspace*{\fill}\\
\hlstd{}\hlstd{\ \ \ \ }\hlstd{Result1\ }\hlopt{=}\hspace*{\fill}\\
\hlstd{}\hlstd{\ \ \ \ \ \ }\hlstd{}\hlkwa{case\ }\hlstd{}\hlopt{\{}\hlstd{Result}\hlopt{,\ }\hlstd{HC}\hlopt{\}\ }\hlstd{}\hlkwa{of}\hspace*{\fill}\\
\hlstd{}\hlstd{\ \ \ \ \ \ \ \ \ }\hlstd{}\hlopt{\{}\hlstd{undefined}\hlopt{,\ }\hlstd{undefined}\hlopt{\}\ {-}$>$}\hspace*{\fill}\\
\hlstd{}\hlstd{\ \ \ \ \ \ \ \ \ \ \ \ }\hlstd{Result}\hlopt{;}\hspace*{\fill}\\
\hlstd{}\hlstd{\ \ \ \ \ \ \ \ \ }\hlstd{}\hlopt{\{}\hlstd{undefined}\hlopt{,\ }\hlstd{\textunderscore }\hlopt{\}\ {-}$>$}\hspace*{\fill}\\
\hlstd{}\hlstd{\ \ \ \ \ \ \ \ \ \ \ \ }\hlstd{}\hlopt{\{}\hlstd{HC}\hlopt{,\ }\hlstd{PosN}\hlopt{\};}\hspace*{\fill}\\
\hlstd{}\hlstd{\ \ \ \ \ \ \ \ \ }\hlstd{}\hlopt{\{\{}\hlstd{candidate}\hlopt{,\ }\hlstd{\textunderscore HCC}\hlopt{,\ }\hlstd{PosC}\hlopt{,\ }\hlstd{\textunderscore PosI}\hlopt{\},\ }\hlstd{\textunderscore }\hlopt{\}\ }\hlstd{}\hlkwa{when\ }\hlstd{PosN\ }\hlopt{$>$\ }\hlstd{PosC\ }\hlopt{{-}$>$}\hspace*{\fill}\\
\hlstd{}\hlstd{\ \ \ \ \ \ \ \ \ \ \ \ }\hlstd{Result}\hlopt{;}\hspace*{\fill}\\
\hlstd{}\hlstd{\ \ \ \ \ \ \ \ \ }\hlstd{}\hlopt{\{\{}\hlstd{candidate}\hlopt{,\ }\hlstd{HCC}\hlopt{,\ }\hlstd{PosC}\hlopt{,\ }\hlstd{PosI}\hlopt{\},\ }\hlstd{undefined}\hlopt{\}\ }\hlstd{}\hlkwa{when\ }\hlstd{PosN\ }\hlopt{=$<$\ }\hlstd{PosI\ }\hlopt{{-}$>$}\hspace*{\fill}\\
\hlstd{}\hlstd{\ \ \ \ \ \ \ \ \ \ \ \ }\hlstd{}\hlopt{\{}\hlstd{candidate}\hlopt{,\ }\hlstd{HCC}\hlopt{,\ }\hlstd{PosC\ }\hlopt{+\ }\hlstd{}\hlnum{1}\hlstd{}\hlopt{,\ }\hlstd{PosN}\hlopt{\};}\hspace*{\fill}\\
\hlstd{}\hlstd{\ \ \ \ \ \ \ \ \ }\hlstd{}\hlopt{\{\{}\hlstd{candidate}\hlopt{,\ }\hlstd{HCC}\hlopt{,\ }\hlstd{PosC}\hlopt{,\ }\hlstd{PosI}\hlopt{\},\ }\hlstd{undefined}\hlopt{\}\ {-}$>$}\hspace*{\fill}\\
\hlstd{}\hlstd{\ \ \ \ \ \ \ \ \ \ \ \ }\hlstd{}\hlopt{\{}\hlstd{candidate}\hlopt{,\ }\hlstd{HCC}\hlopt{,\ }\hlstd{PosC\ }\hlopt{+\ }\hlstd{}\hlnum{1}\hlstd{}\hlopt{,\ }\hlstd{PosI}\hlopt{\};}\hspace*{\fill}\\
\hlstd{}\hlstd{\ \ \ \ \ \ \ \ \ }\hlstd{}\hlopt{\{\{}\hlstd{candidate}\hlopt{,\ }\hlstd{\textunderscore HCC}\hlopt{,\ }\hlstd{\textunderscore PosC}\hlopt{,\ }\hlstd{PosI}\hlopt{\},\ }\hlstd{\textunderscore }\hlopt{\}\ }\hlstd{}\hlkwa{when\ }\hlstd{PosN\ }\hlopt{=$<$\ }\hlstd{PosI\ }\hlopt{{-}$>$}\hspace*{\fill}\\
\hlstd{}\hlstd{\ \ \ \ \ \ \ \ \ \ \ \ }\hlstd{}\hlopt{\{}\hlstd{HC}\hlopt{,\ }\hlstd{PosN}\hlopt{\};}\hspace*{\fill}\\
\hlstd{}\hlstd{\ \ \ \ \ \ \ \ \ }\hlstd{}\hlopt{\{\{}\hlstd{candidate}\hlopt{,\ }\hlstd{\textunderscore HCC}\hlopt{,\ }\hlstd{\textunderscore PosC}\hlopt{,\ }\hlstd{PosI}\hlopt{\},\ }\hlstd{\textunderscore }\hlopt{\}\ {-}$>$}\hspace*{\fill}\\
\hlstd{}\hlstd{\ \ \ \ \ \ \ \ \ \ \ \ }\hlstd{}\hlopt{\{}\hlstd{candidate}\hlopt{,\ }\hlstd{HC}\hlopt{,\ }\hlstd{PosN}\hlopt{,\ }\hlstd{PosI}\hlopt{\};}\hspace*{\fill}\\
\hlstd{}\hlstd{\ \ \ \ \ \ \ \ \ }\hlstd{}\hlopt{\{\{}\hlstd{\textunderscore HCC}\hlopt{,\ }\hlstd{PosC}\hlopt{\},\ }\hlstd{\textunderscore }\hlopt{\}\ }\hlstd{}\hlkwa{when\ }\hlstd{PosN\ }\hlopt{$>$\ }\hlstd{PosC\ }\hlopt{{-}$>$}\hspace*{\fill}\\
\hlstd{}\hlstd{\ \ \ \ \ \ \ \ \ \ \ \ }\hlstd{Result}\hlopt{;}\hspace*{\fill}\\
\hlstd{}\hlstd{\ \ \ \ \ \ \ \ \ }\hlstd{}\hlopt{\{\{}\hlstd{HCC}\hlopt{,\ }\hlstd{PosC}\hlopt{\},\ }\hlstd{undefined}\hlopt{\}\ {-}$>$}\hspace*{\fill}\\
\hlstd{}\hlstd{\ \ \ \ \ \ \ \ \ \ \ \ }\hlstd{}\hlopt{\{}\hlstd{candidate}\hlopt{,\ }\hlstd{HCC}\hlopt{,\ }\hlstd{PosC\ }\hlopt{+\ }\hlstd{}\hlnum{1}\hlstd{}\hlopt{,\ }\hlstd{PosN}\hlopt{\};}\hspace*{\fill}\\
\hlstd{}\hlstd{\ \ \ \ \ \ \ \ \ }\hlstd{}\hlopt{\{\{}\hlstd{\textunderscore HCC}\hlopt{,\ }\hlstd{\textunderscore PosC}\hlopt{\},\ }\hlstd{\textunderscore }\hlopt{\}\ {-}$>$}\hspace*{\fill}\\
\hlstd{}\hlstd{\ \ \ \ \ \ \ \ \ \ \ \ }\hlstd{}\hlopt{\{}\hlstd{HC}\hlopt{,\ }\hlstd{PosN}\hlopt{\}}\hspace*{\fill}\\
\hlstd{}\hlstd{\ \ \ \ \ \ }\hlstd{}\hlkwa{end}\hlstd{}\hlopt{,}\hspace*{\fill}\\
\hlstd{}\hlstd{\ \ \ \ }\hlstd{}\hlkwd{consistent\textunderscore hash}\hlstd{}\hlopt{(}\hlstd{Result1}\hlopt{,\ }\hlstd{Key\ div\ CurrentBase}\hlopt{,\ }\hlstd{HashCodes}\hlopt{,}\hspace*{\fill}\\
\hlstd{}\hlstd{\ \ \ \ \ \ \ \ \ \ \ \ \ \ \ \ \ \ \ \ }\hlstd{CurrentBase\ }\hlopt{+\ }\hlstd{}\hlnum{1}\hlstd{}\hlopt{).}\hspace*{\fill}\\
\mbox{}
\normalfont
\normalsize
\caption{Improving \fref{fig:code:new-one-result} by accommodating removed hash codes}
\label{fig:code:new-one-result-marker}
\end{figure}

Whilst there are ten combinations of existing-result and new hash code
to consider in this code, the algorithmic complexity is no worse, and
so when just a single result is required, the cost of the function is
$\BigO{|C|}$. Whilst this is still worse than for the classic
algorithm, for the smaller sizes of $|C|$ that our algorithm is best
suited for, this is unlikely to preclude use of our
algorithm. Equally, in cases where there is a very high churn rate of
hash codes being added and removed, the lower cost of these operations
in our algorithm may favour it over the classic algorithm.

\section{Evaluation}

As mentioned earlier, whilst our algorithm achieves perfect uniformity
along with the remapping and determinism properties, the trade-off is
higher average cost ($\BigO{|C| \cdot \log_2(|C|)}$ (or $\BigO{|C|}$
if a single element is returned rather than an entire permutation)
versus $\BigO{\log_2(|C|)}$ for the classic algorithm) and rapid
consumption of the entropy of the key. If more hash codes are used
than can be supported by the entropy of the key then the consequence
is certain permutations will never be reached and thus certain hash
codes may never appear at the front of resulting
permutations. However, our algorithm can dynamically construct the
result as the key is consumed, thus avoiding building the entire tree,
saving memory. This is possible because the contents of each child
node are determined by just the remainder of the key and the node's
hash code itself. By contrast in the classic algorithm, the contents
of each child node are determined randomly by the placement of hash
code points. This means at a minimum, all the points of every hash
code must be held in memory for the classic algorithm.

Returning a permutation rather than a single result is in practice
very useful, and has several interpretations depending on the
application. For example, if the application is a distributed
key-value store then the permutation would indicate an ordering of
machines to try: in the case of a read operation you might choose to
issue reads to the first few machines from the permutation, either to
check that they all have the same value, or because due to transient
load imbalances, one may reply more quickly than the others. For a
write operation, the client application may well indicate that it only
considers a write {\em completed} once it has been synchronously
written to at least $N$ machines; again, the first $N$ elements from
the permutation indicate exactly to which machines to issue
synchronous writes.

In such key-value stores, it is very often the case that certain keys
are much more frequently accessed than others. This might be due to a
particularly popular URL; the effect of {\em ``being slash-dotted''}
or {\em ``going viral''}. In these scenarios, a single key can
substantially skew loading across a cluster of machines. Here again,
returning a permutation from the consistent hash function can be
advantageous: if each element of the permutation is a particular
machine in your distributed key-value store and loading information
per machine is available, the client may well be able to filter out
particularly heavily loaded machines and still access the required
information promptly. In an eventually consistent scenario with writes
as well as reads occurring, this could result in the serving of stale
data, but the trade-off would be better load balancing and improved
latencies.

Performance comparisons in general of the classic algorithm and our
new algorithm are of limited value as they will inevitably reflect
both the suitability of each algorithm to the artificial conditions of
the benchmark, and the relative amounts of effort to optimise each
implementation.

\section{Conclusion}

Consistent Hashing is a widely used and important technique,
applicable to many applications. Hopefully this work provides a more
detailed understanding as to how it can be achieved and what the
trade-offs involved are.

We have shown how the classic algorithm relies on random distributions
to approximately maintain the uniformity property. We then examined
how, given the way in which the circle achieves the remapping
property, universal cycles of shorthand permutations may be used to
precisely achieve and maintain the uniformity property in light of
removals of hash codes but that adding new hash codes is
non-obvious. Finally, by abandoning the use of a circle, we presented
our new algorithm which also relies on permutations but makes the
addition of hash codes simple. We have discussed potential
implementation strategies and average performance of these algorithms
and shown that a cost of achieving perfect uniformity and remapping is
in the factorial relationship between the number of hash codes and the
key size.

\bibliographystyle{alpha}
\bibliography{../../library}

\end{document}